\documentclass{epsconf}

\usepackage[british]{babel}
\usepackage[english=british]{csquotes}
\usepackage{graphicx}
\usepackage{wrapfig}
\usepackage{amsmath}
\usepackage{bm}
\usepackage{siunitx}
\usepackage{cleveref}
\usepackage[nolist]{acronym}
\usepackage[format=plain,labelfont=bf,font=footnotesize]{caption}
\usepackage{setspace}
\usepackage{hyperref}
\usepackage{caption}
\usepackage{subcaption}

\graphicspath{{figures/}}

\title{
  Kinetic Modeling of Seeded Nitrogen in an ITER Baseline Scenario
}
\author{%
  \underline{F.~Schluck}$^1$
}
\institute{
  $^1$Forschungszentrum J\"ulich GmbH, Institut f\"ur Energie- und 
  Klimaforschung -- Plasmaphysik, Partner of the Trilateral Euregio Cluster (TEC), 52425 J\"ulich, Germany
}
\begin{document}
\maketitle

{\footnotesize\noindent Keywords: Nuclear Fusion, Magnetic Confinement, Theory and Modeling, ITER, Nitrogen Seeding, EIRENE\\
J\"ulich, \today}

\section*{Abstract}
ITER as the next-level fusion device is intended to reliably produce more fusion power than required for sustainably heating its plasma. Modeling has been an essential part of the ITER design and for planning of future experimental campaigns. In a tokamak or stellarator plasma discharge, impurities play a significant role, especially in the edge region. Residual gases, eroded wall material, or even intentionally seeded gases all heavily influence the confinement and, thus, the overall fusion performance. Nitrogen is such a gas envisaged to be seeded into a discharge plasma. By modeling the impurities kinetically using the full three-dimensional Monte-Carlo code package EMC3-EIRENE, we analyze the distribution of nitrogen charge-state resolved in a seeded ITER baseline scenario and draw conclusions for the hydrogen background plasma density. Lastly, we compare the influence of a more refined kinetic ion transport in EIRENE including additional physical effects on the impurity density.

\section{Introduction}
By completing the ITER tokamak \cite{iter}, magnetic confinement plasma physics will reach a next step on its way to fusion energy. As the completion will still take some years, modeling via numerical simulation is the favored tool to preplan experiments and estimate ideal machine settings for optimized plasma discharge parameters \cite{model1,model2,model3,model4}. To intentionally seed an impurity of different species, density, and energy is one tunability option in a fusion device. Ideally, such a trace gas allows for a distinct energy loss of the main plasma species \cite{seeding1}, for setting the heat deposition on the target \cite{seeding2}, for reaching detachment \cite{seeding3}, and, overall, for improving the confinement \cite{seeding4}.

The three-dimensional Monte-Carlo code EIRENE \cite{eirene1,eirene2}, usually used for kinetically modeling neutral particle transport in the plasma edge region, is a reliable and widely used simulation tool for mainly three reasons: it is coupled to broad databases of atomic, molecular, and other reactive data \cite{hydkin}, it is applied to realistic large-scale geometries \cite{iterdivertor}, and it is combined with all major European plasma edge codes \cite{fluid1,fluid2,fluid3}. The three-dimensional edge Monte-Carlo code EMC3 \cite{emc3} is also coupled to EIRENE into the code package EMC3-EIRENE, allowing to study plasma discharges in full 3D. 

Some charged particles might have a short lifetime compared to their collisionality $\tau_\text{life}\ll\tau_\text{coll}$. Handling of such ions requires a kinetic treatment, as assuming local thermodynamic equilibrium is deceptive. Recently, the capability of describing ions kinetically via EIRENE in EMC3-EIRENE context has been extended \cite{eps18,iaea,ctpp} by first order drift effects, anomalous cross-field transport, and regarding the mirror force, while beforehand only field-line tracing and energy relaxation have been included.

We take an ITER hydrogen plasma baseline scenario model from the 2008 database \cite{iter} and perturb it by installing a nitrogen density gas puff, where we perform a charge state resolved kinetic modeling of nitrogen and analyze the seeding's impact on the main plasma. We compare the different physics model in EIRENE, namely the ``classic'' one including only field-line tracing and energy relaxation due to the background conditions to the ``enhanced'' one, adding drifts, diffusion, and mirror force. This is an exemplary case study which can be easily expanded to study different or additional impurity species.

The paper is organized as follows. In Sec. \ref{model} we introduce the simulation setup of the EMC3-EIRENE runs and explain how a kinetically simulated impurity species influences the main plasma. Afterwards, we present results on the aforementioned perturbed ITER baseline modeling in Sec. \ref{results}. Sec. \ref{comparison} focuses on the comparison between the different fidelity in the kinetic ion transport of EIRENE, while we will close in Sec. \ref{summary} by summarizing the results and drawing conclusions for future endeavors.

\section{EMC3-EIRENE Simulation Setup}\label{model}
EMC3 and EIRENE are coupled iteratively, meaning that at first the plasma background equilibrates in the magnetic structure. Afterwards, EIRENE particles are started on that background, probabilities for collisional events depend on the current plasma density and temperature. Neutrals of the same atomic species as the main plasma can change their character, which means that e.g. in the case studied in this manuscript the kinetically treated neutral molecule in EIRENE $H_2$ might switch towards a main plasma species fluid parcel of $H^+$ in EMC3, which eventually could become an again kinetically treated minority ion $H_2^+$ in EIRENE. If an EIRENE particle is not of the main plasma species, interaction from EIRENE towards EMC3 happens only via the electron momentum balance and temperature equation. This iterative coupling between EMC3 and EIRENE is performed until the solution converges. For a more detailed explanation on the underlying physics equations, see \cite{emc3,eirene1,eirene2,ctpp}.

\begin{figure}[h]
\includegraphics[width=\linewidth]{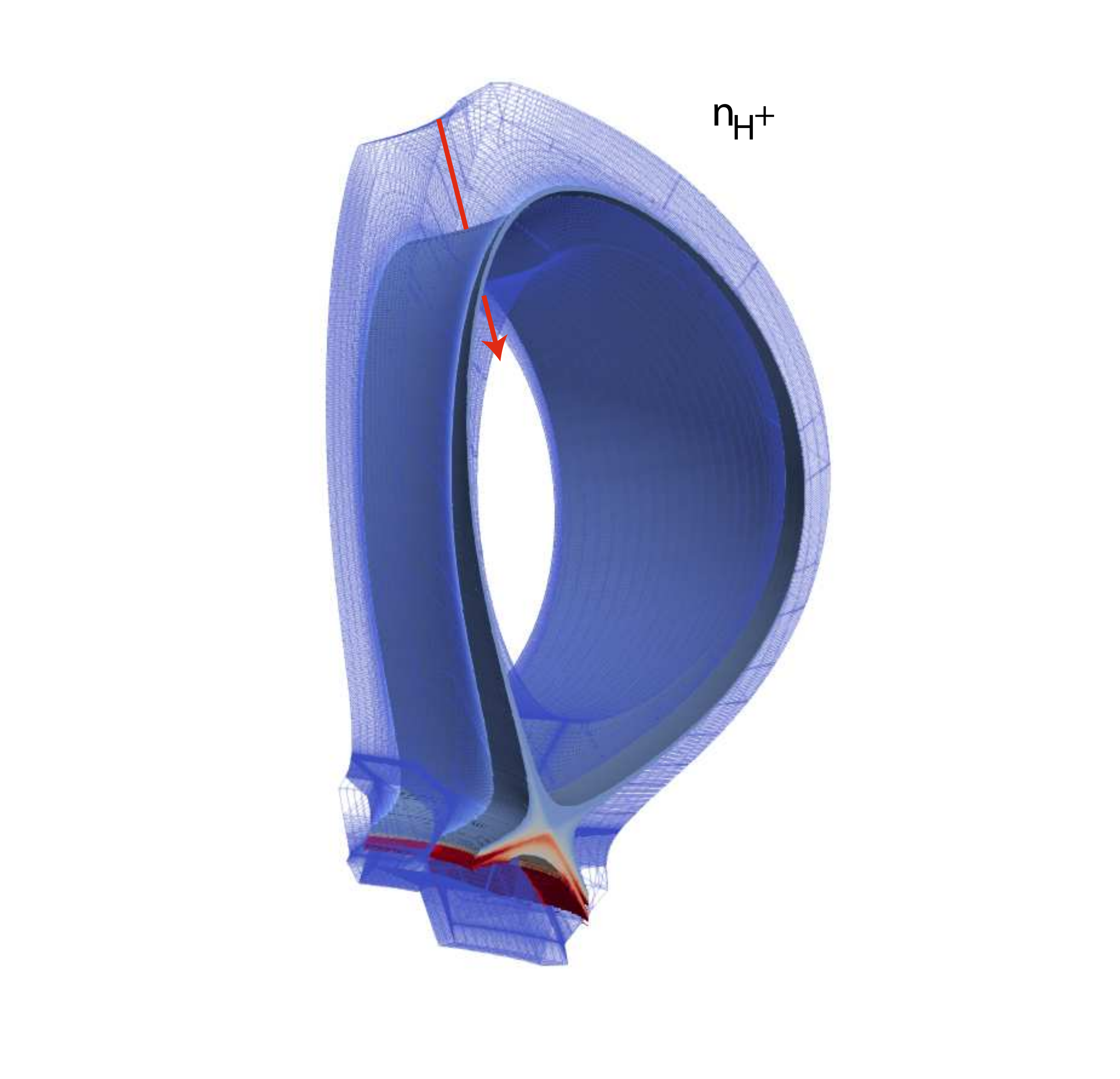}
\caption{Schematic simulation setup of the ITER baseline case as described in the text. Shown is the simulation volume, the EMC3 brick structure, and, exemplarily, the proton density $n_{H^+}$, the main plasma species.}
\label{fig:scheme}
\end{figure}

In Fig. \ref{fig:scheme} one finds the simulation area, which consists of an axisymmetric toroidal section of 40° of the ITER tokamak with adequate boundary conditions. Examplarily, the main plasma species density $n_{H^+}$ is plotted. 

The simulated ITER case 2297 $\Psi_N=0.83$ is a hydrogen plasma case from the 2008 baseline scenario \cite{iter}, where EIRENE is handling the hydrogen recycling at the target plates plus a constant molecular hydrogen gas puff in order to establish semi-detached plasma conditions. The different reactions for the hydrogen species $H_2$, $H$, and $H_2^+$ in EIRENE comprise electron impact ionization, recombination, charge exchange, elastic collision, and dissociation.

We add a monoenergetic molecular nitrogen gas puff with $E_{N_2}=0.026$ eV at the hydrogen puffing location, cf. red arrow in Fig. \ref{fig:scheme}, with a flux of $\Phi=3.2\times10^{15}$ s$^{-1}$ and a total number of puffed-in nitrogen of one tenth of the amount of puffed-in hydrogen. It is possible for the molecular nitrogen to dissociate in three different ways \cite{amjuel1},
\begin{align}
 N_2 + e &\rightarrow 2N + e + \Delta E_\text{el}, & \Delta E_\text{el} &= 9.7527~\text{eV},\\
 N_2 + e &\rightarrow N_2^+ + 2e + \Delta E_\text{el}, & \Delta E_\text{el} &= 15.581~\text{eV},\\
 N_2 + e &\rightarrow N^+ + N + 2e + \Delta E_\text{el}, & \Delta E_\text{el} &= 24.34~\text{eV},
\end{align}
while the molecular ion has two implemented reactions \cite{amjuel2}, namely
\begin{align}
 N_2^+ + e &\rightarrow 2N^+ + 2e + \Delta E_\text{el}, & \Delta E_\text{el} &= 31.2~\text{eV},\\
 N_2^+ + e &\rightarrow N + N^+ + e + \Delta E_\text{el}, & \Delta E_\text{el} &= 8.4~\text{eV}.
\end{align}
For the atomic ions we add the possibility to ionize up and recombine down in their charge state level, following the data provided in the ADAS database \cite{adas}. $N^{7+}$ only has the opportunity to recombine.

The simulation area splits into three zones: the confined core region, where field lines are closed, the scrape-off layer, where field lines enter the wall segments, and a private flux area below the divertor.

\section{Kinetic Modeling of Nitrogen}\label{results}
\begin{figure}[h]\centering
\begin{minipage}{72mm}
\includegraphics[width=\linewidth]{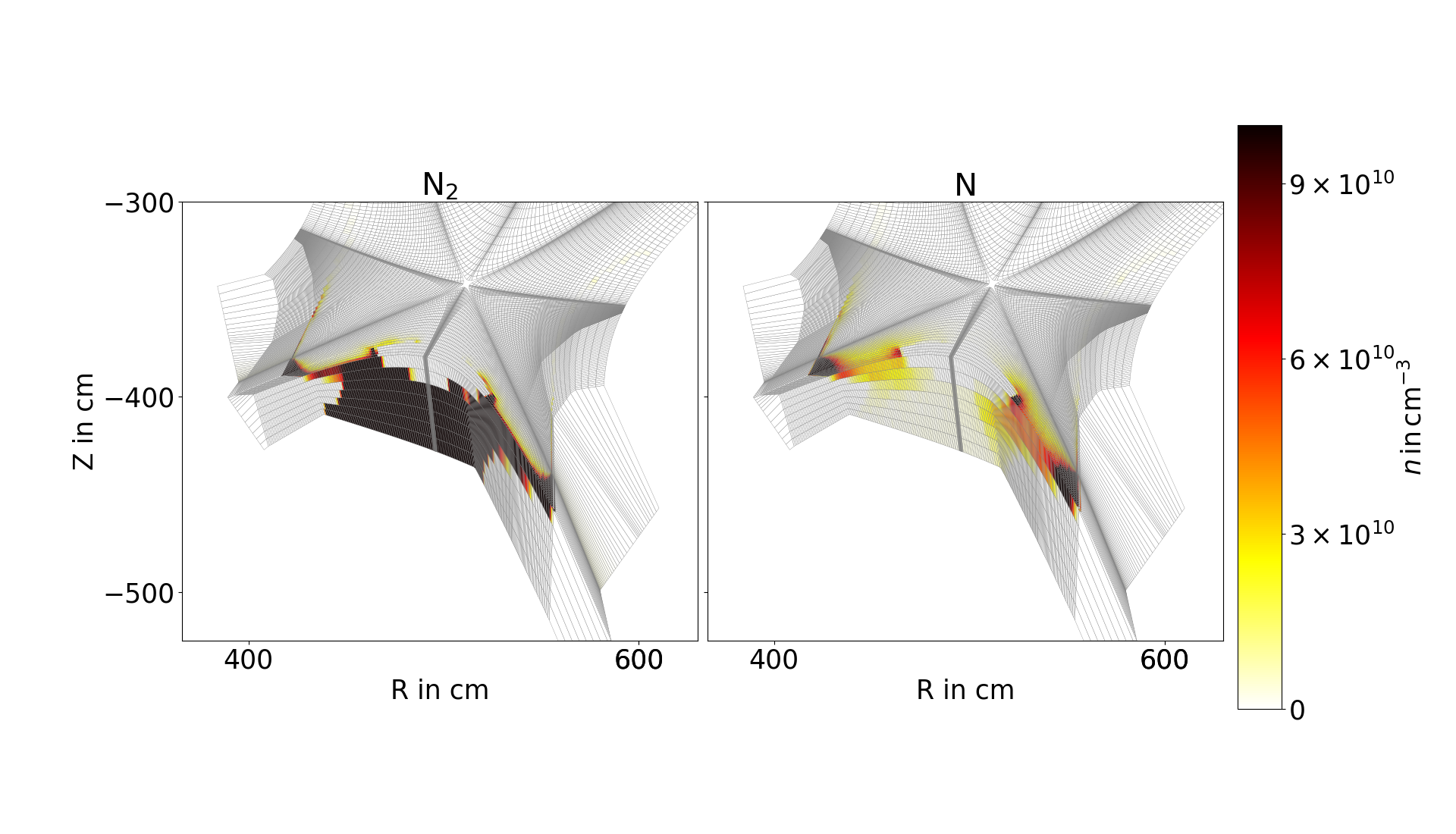}
\caption{Toroidally averaged density distribution of $N_2$ and $N$ in the divertor region.}
\label{fig:1}
\end{minipage}
\hfil
\begin{minipage}{72mm}
\includegraphics[width=\linewidth]{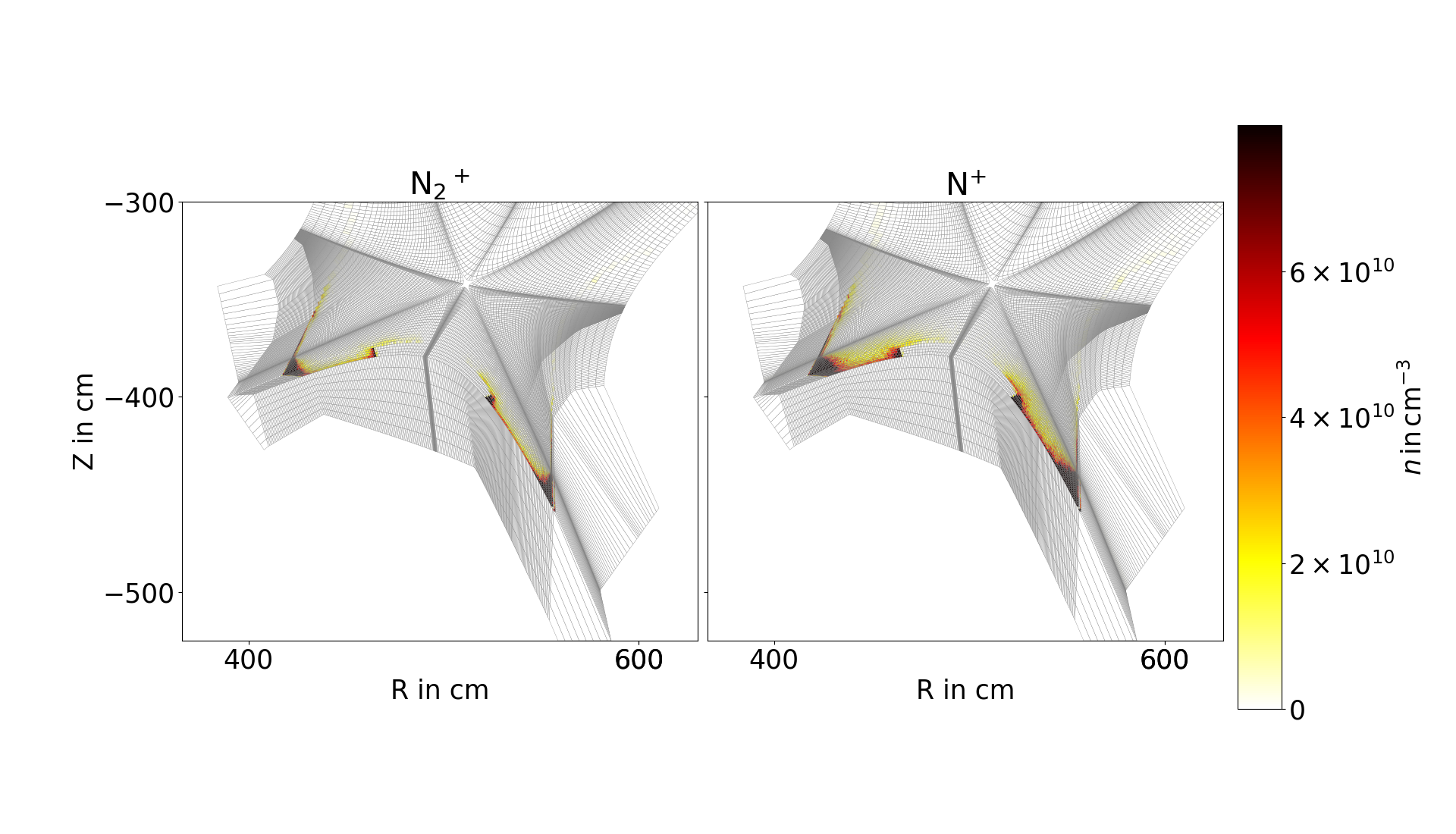}
\caption{Toroidally averaged density distribution of $N_2^{+}$ and $N^{+}$ in the divertor region.}
\label{fig:2}
\end{minipage}\\

\begin{minipage}{72mm}
\includegraphics[width=\linewidth]{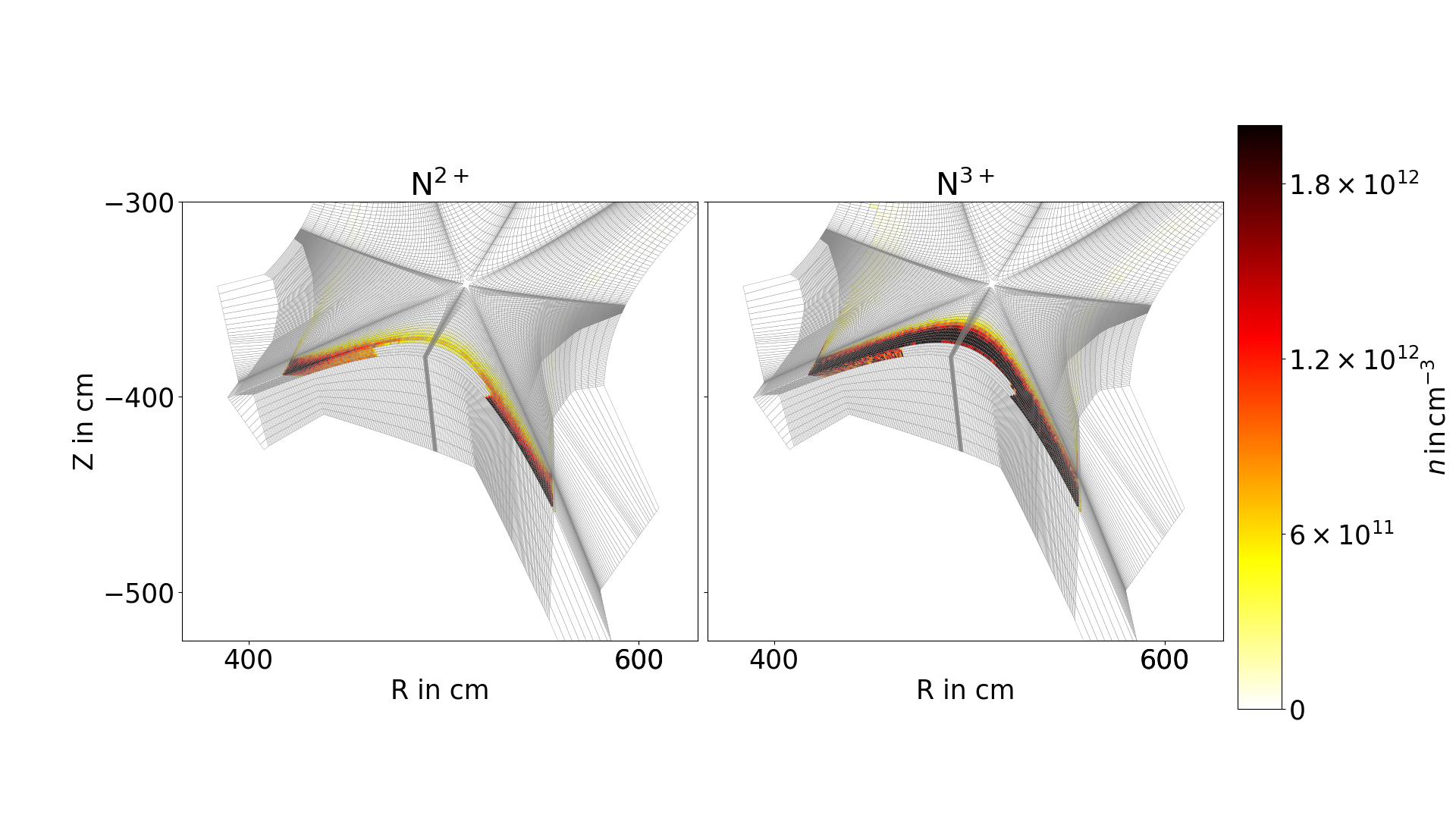}
\caption{Toroidally averaged density distribution of $N^{2+}$ and $N^{3+}$ in the divertor region.}
\label{fig:3}
\end{minipage}
\hfil
\begin{minipage}{72mm}
\includegraphics[width=\linewidth]{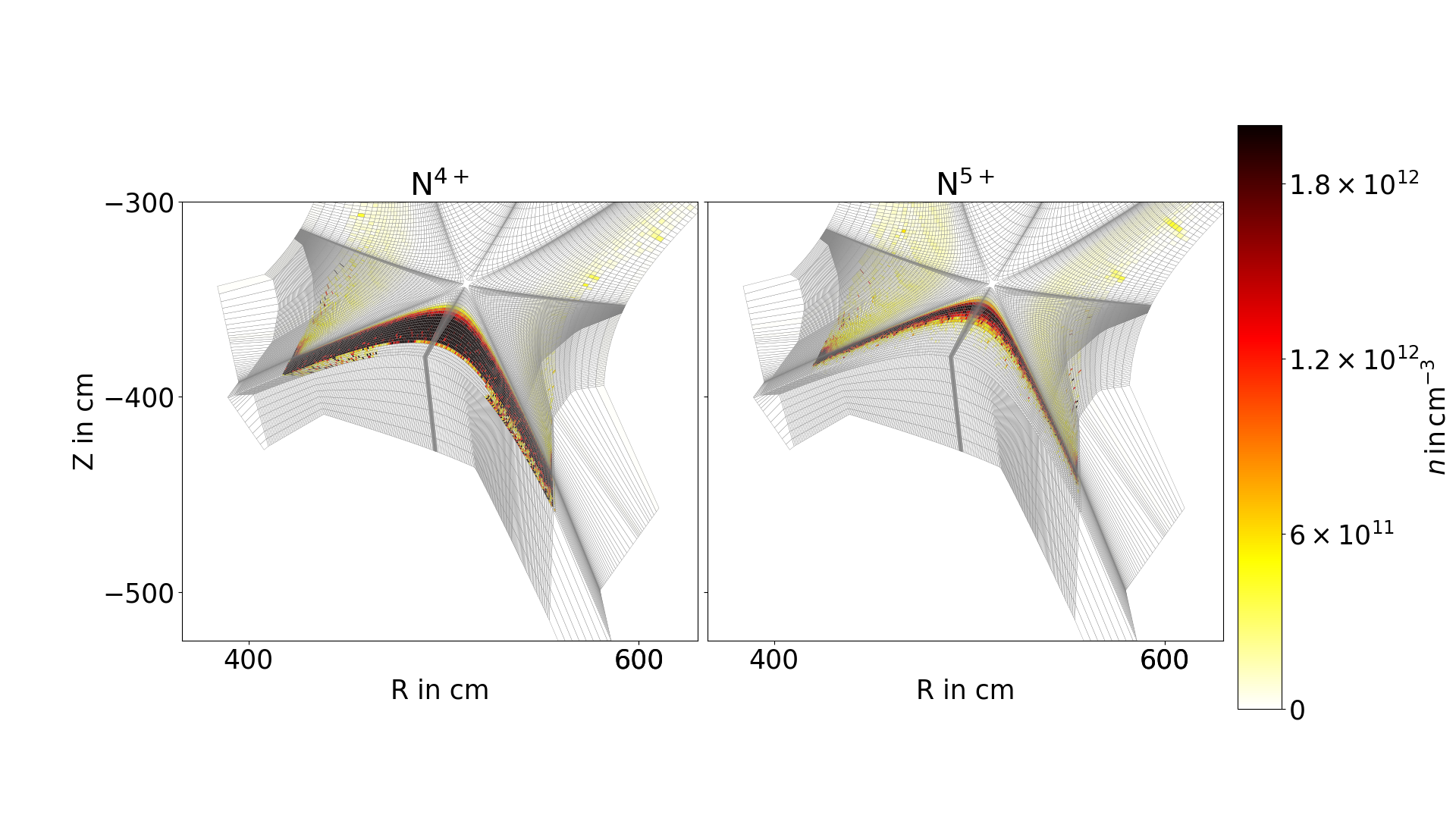}
\caption{Toroidally averaged density distribution of $N^{4+}$ and $N^{5+}$ in the divertor region.}
\label{fig:4}
\end{minipage}\\

\begin{minipage}{72mm}
\includegraphics[width=\linewidth]{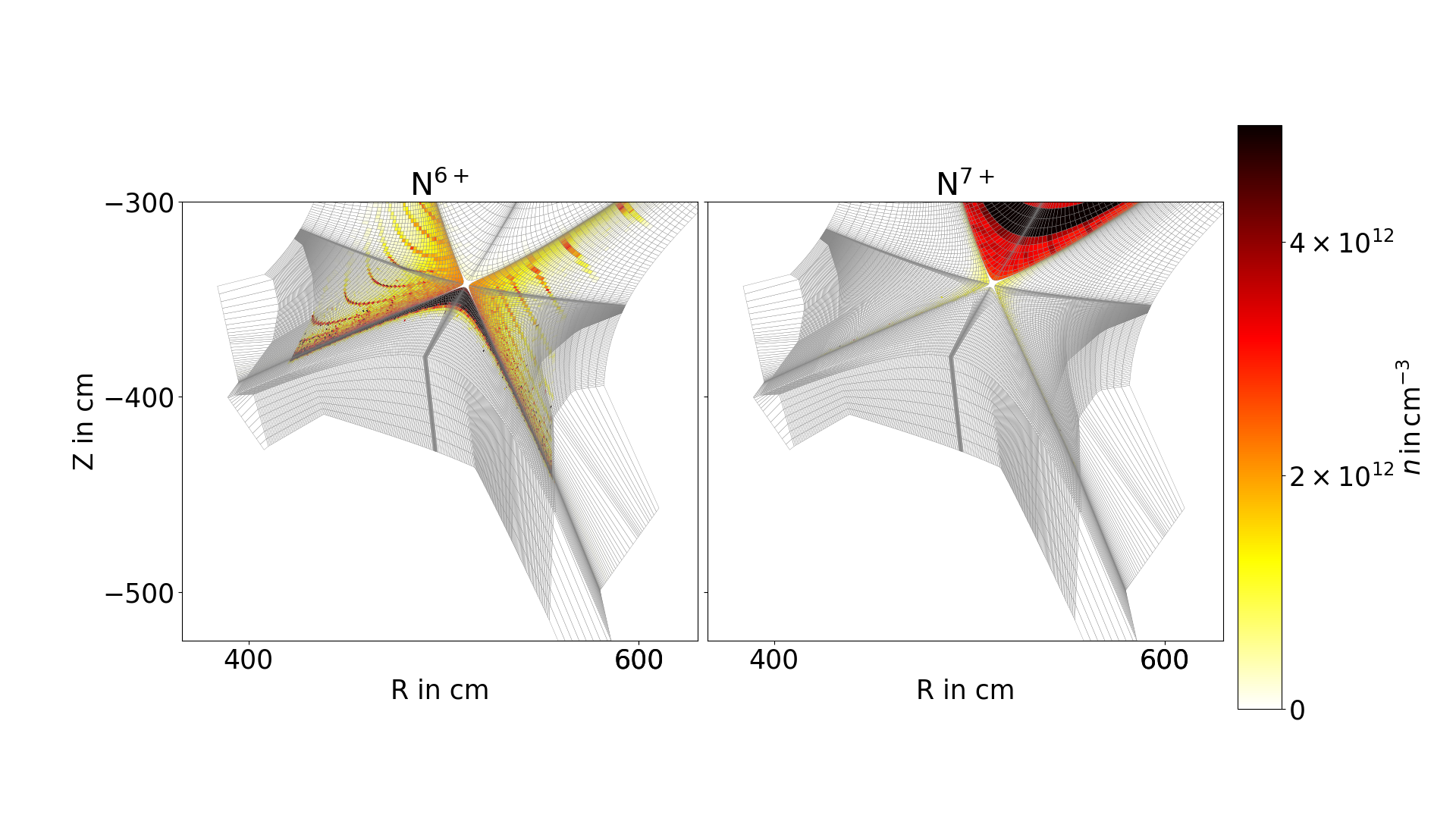}
\caption{Toroidally averaged density distribution of $N^{6+}$ and $N^{7+}$ in the divertor region.}
\label{fig:5}
\end{minipage}
\end{figure}

We start investigating the density distribution of the different nitrogen charge states in the divertor region. Figs. \ref{fig:1}--\ref{fig:5} show the densities of $N_2$, $N$, $N_2^+$, $N^+$, $N^{2+}$, $N^{3+}$, $N^{4+}$, $N^{5+}$, $N^{6+}$, and $N^{7+}$.

In Fig. \ref{fig:1} one notices the molecular nitrogen $N_2$ being present mainly in the private flux region below the divertor dome. As it is not charged, it can reach this area, and once it leaves the private flux area it gets ionized due to the background conditions, thus leaving no density behind in the scrape-off layer. For atomic nitrogen $N$, one notices a relatively large density at the divertor target plates which is due to neutralizing when catching a surface electron during a reflection. With peak densities of less than $10^{11}$ cm$^{-3}$, both the density for $N_2$ and $N$ are relatively low.

One finds even lower densities for the molecular ion $N_2^+$ and the singlely charged nitrogen atom $N^+$ in Fig. \ref{fig:2}. Again, as for $N$, one finds the density peaked close to the target plates at the divertor, implying that the background plasma is energetic enough to immediately ionize both $N_2^+$ and $N^+$ further up, as is also clear from the relatively low densities around $7\times10^{10}$ cm$^{-3}$. 

\begin{figure}[h]\centering
\begin{minipage}{72mm}
\includegraphics[width=\linewidth]{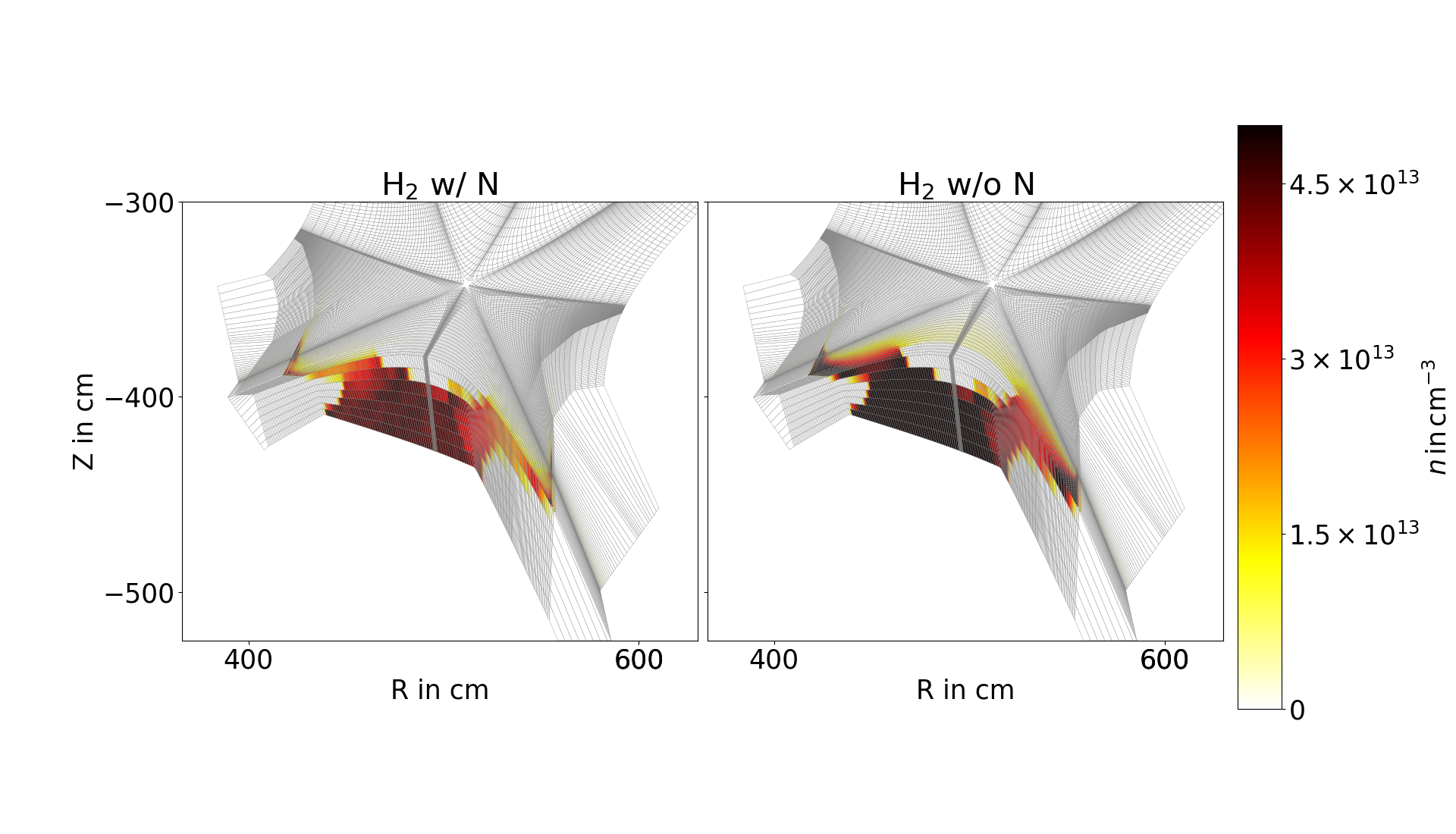}
\caption{Toroidally averaged density distribution of molecular hydrogen $H_2$, comparing scenarios with nitrogen seeding (left) and without (right).}
\label{fig:6}
\end{minipage}
\hfil
\begin{minipage}{72mm}
\includegraphics[width=\linewidth]{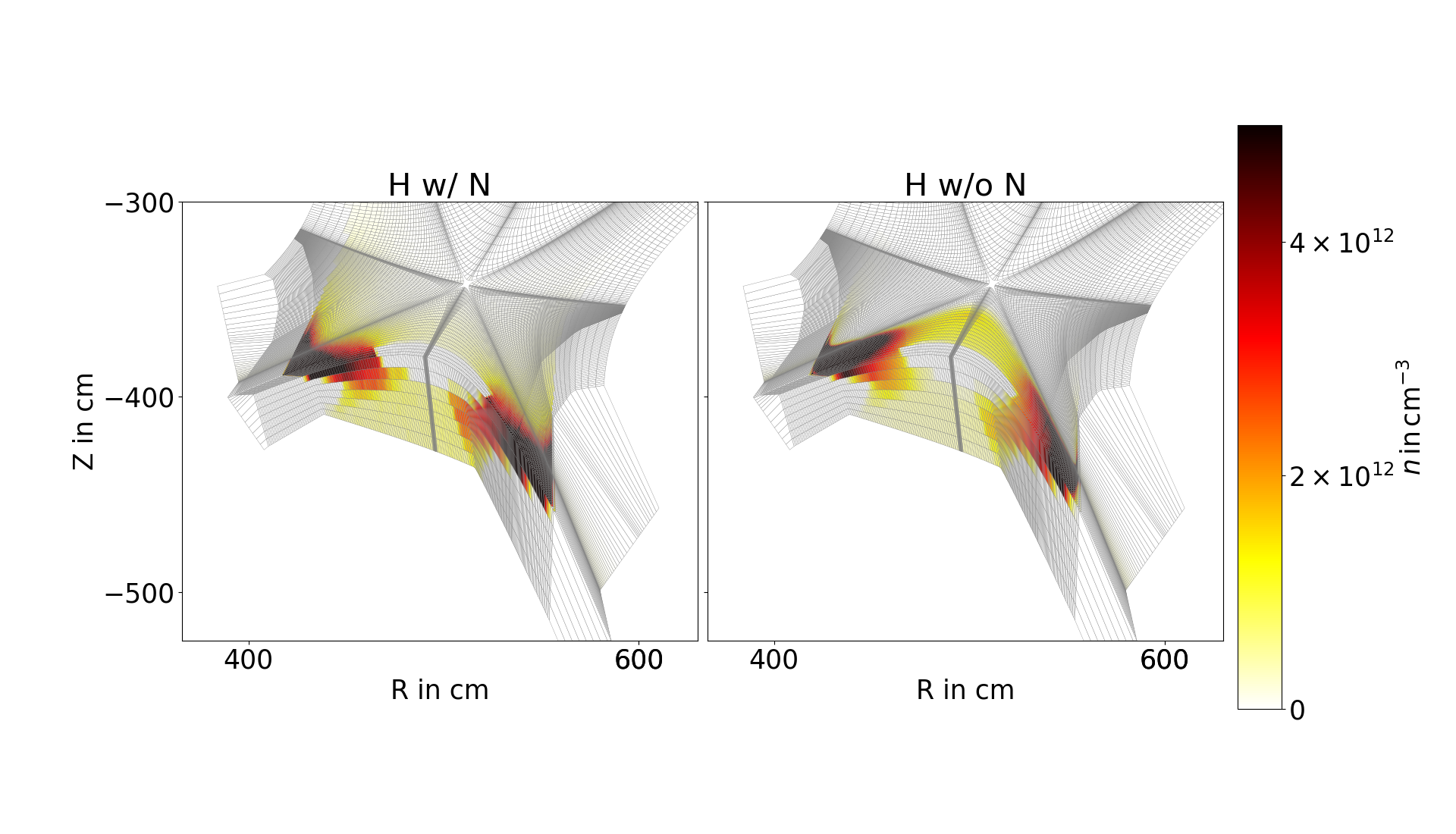}
\caption{Toroidally averaged density distribution of atomic hydrogen $H$, comparing scenarios with nitrogen seeding (left) and without (right).}
\label{fig:7}
\end{minipage}\\

\begin{minipage}{72mm}
\includegraphics[width=\linewidth]{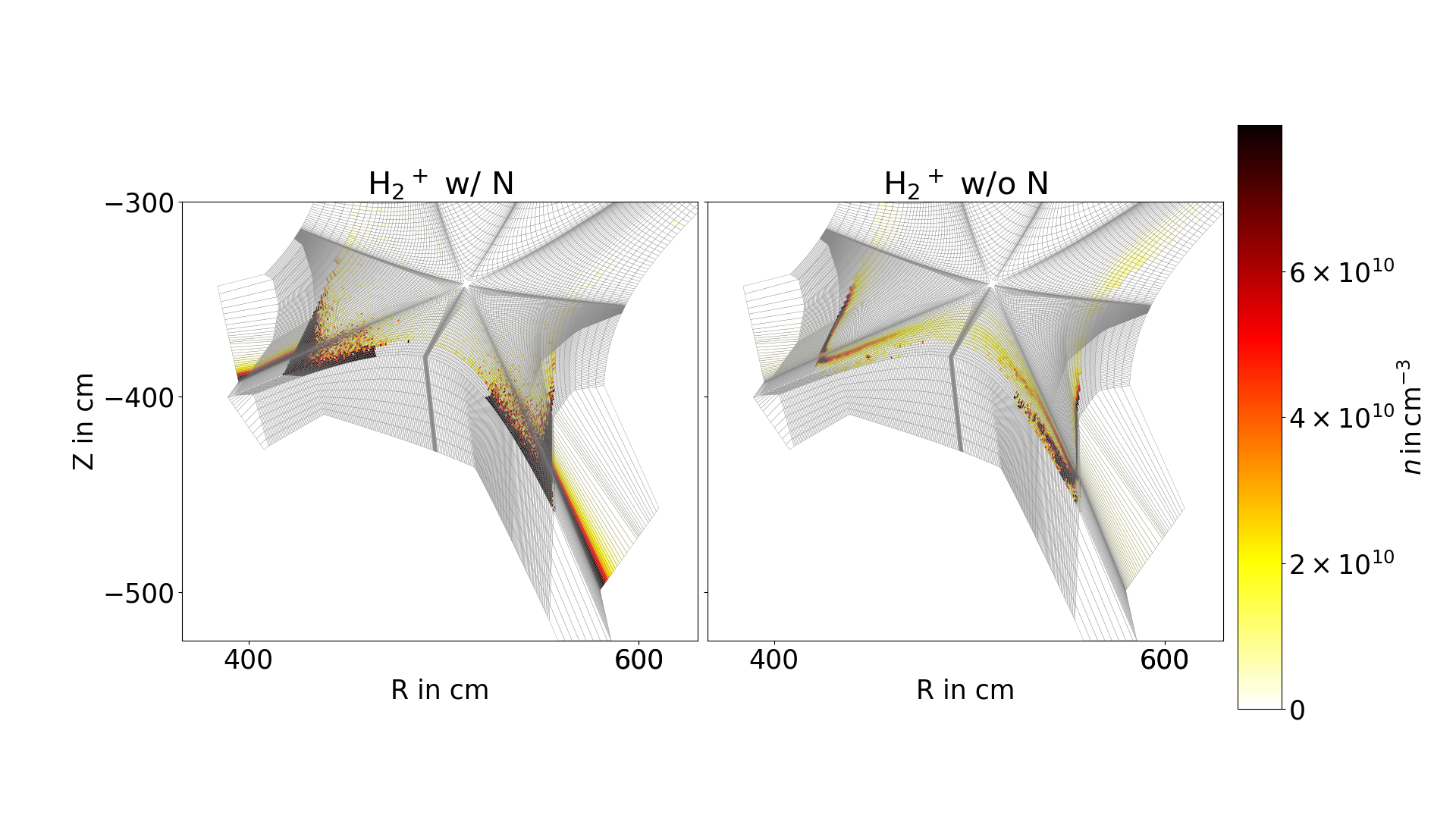}
\caption{Toroidally averaged density distribution of molecular hydrogen ions $H_2^+$, comparing scenarios with nitrogen seeding (left) and without (right).}
\label{fig:8}
\end{minipage}
\hfil
\begin{minipage}{72mm}
\includegraphics[width=\linewidth]{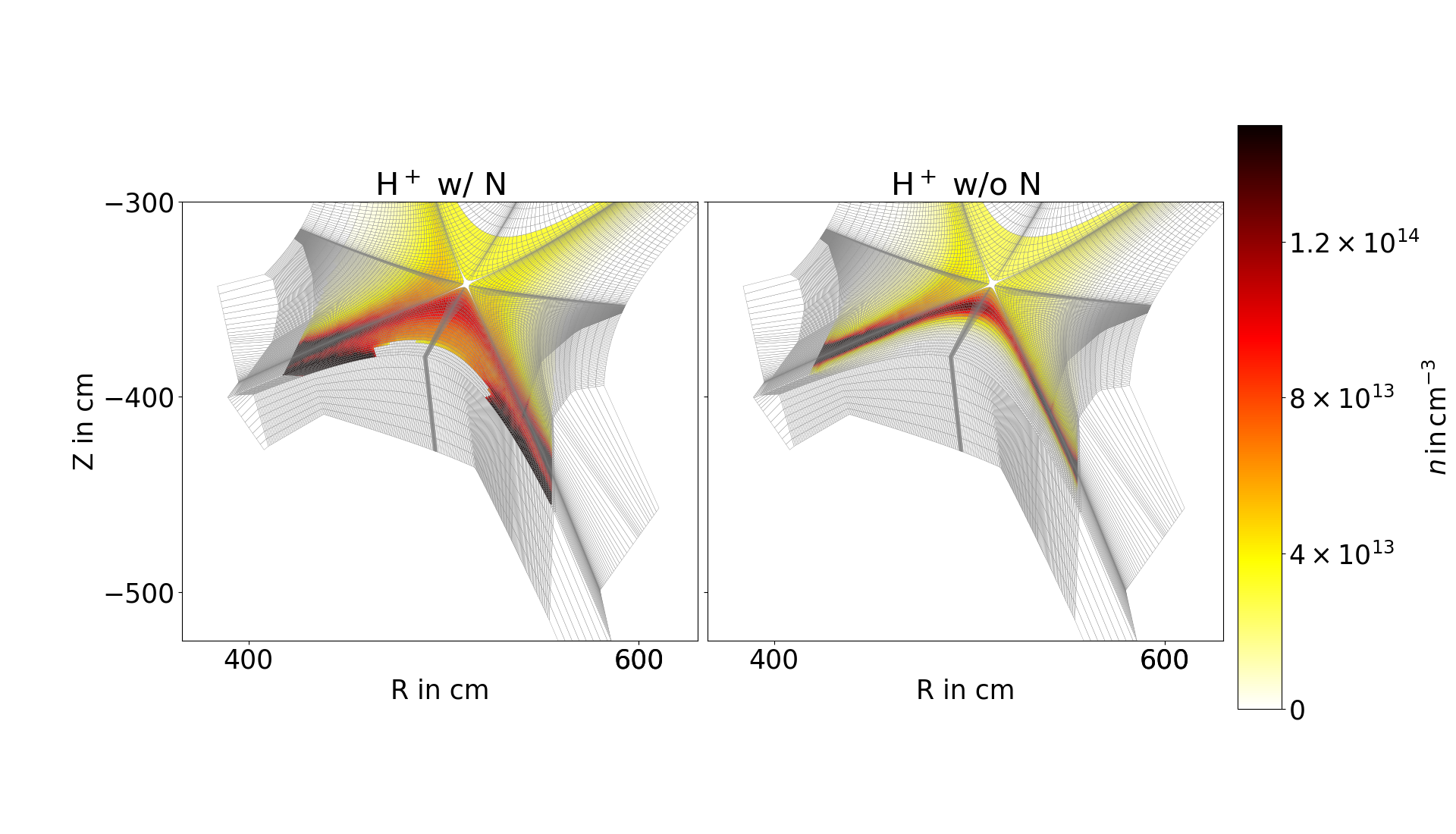}
\caption{Toroidally averaged density distribution of the main plasma species $H^+$, comparing scenarios with nitrogen seeding (left) and without (right).}
\label{fig:9}
\end{minipage}
\end{figure}

For $N^{2+}$ to $N^{6+}$ we find a certain trend (cf. Figs. \ref{fig:3}--\ref{fig:5}), the density is steadily increasing and the distribution peak is approaching the separatrix, as for $N^{2+}$ we find the maximum close at the divertor we eventually find $N^{6+}$ being peaked directly at the separatrix. This is again due to the radial increase of density, temperature, and, ultimately, energy of the main plasma, opening up for higher ionization levels towards the center. For the density of $N^{7+}$, as can be found in Fig. \ref{fig:5} on the right-hand-side, one finds the major part to be distributed in the confined area. 
Next, we analyze the effect the nitrogen gas puff has on the main plasma by comparing the densities for molecular hydrogen $H_2$, atomic hydrogen $H$, molecular ionic hydrogen $H_2^+$, which is the minority ion, and protons $H^+$ in Figs. \ref{fig:6}--\ref{fig:9}. Note that the first three species are handled by EIRENE, while the latter is the only species being dealt with on EMC3 side. These figures are structured in the same way; on the left-hand-side we show the density distribution with the nitrogen gas puff turned on, on the right-hand-side there is no puffed in nitrogen. We remark that the integrated absolute amount of hydrogen in the vessel is the same in both cases.

For the molecular density, which is presented in Fig. \ref{fig:6}, we notice a similar distribution, mainly in the private flux area. If nitrogen is puffed, however, that density is reduced by a factor of roughly $1.5$. The atomic density exhibits a comparable effect (cf. Fig. \ref{fig:7}), the density peaks close to the divertor targets and is slightly redistributed and reduced in case of nitrogen puffing. The minority ion $H_2^+$ (Fig. \ref{fig:8}) shows strong differences in case of a nitrogen gas puff being applied, however, we remark that the absolute density is significantly lower in this case at around $\sim10^{10}$ cm$^{-3}$. Now, most striking and most significant is the change in the plasma main species $H^+$ in Fig. \ref{fig:9}. While in case of no nitrogen being puffed, we find a peaked density at around $n_{H^+}\approx10^{14}$ cm$^{-3}$ in the divertor region, almost at the separatrix. In case of nitrogen puffing, that rather distinct peaking at the separatrix broadens up widely into the divertor region, plus a significant increase of the overall amount of protons, which is clearly visible by a relatively large density of $n_{H^+}>1.5\times10^{14}$ cm$^{-3}$ being directly applied at the divertor target plates.
\section{Enhanced Kinetic Ion Transport}\label{comparison}
In this section we focus on the impact of the recently introduced changes in the physical model of the kinetic ion transport in EIRENE \cite{ctpp}. In this publication, the enhancements of first-order drift effects $\nabla B$- and curvature-drift, magnetic mirror force, and anomalous cross-field diffusion accounting for turbulence effects have been introduced. 

\begin{figure}[h]
\includegraphics[width=\linewidth]{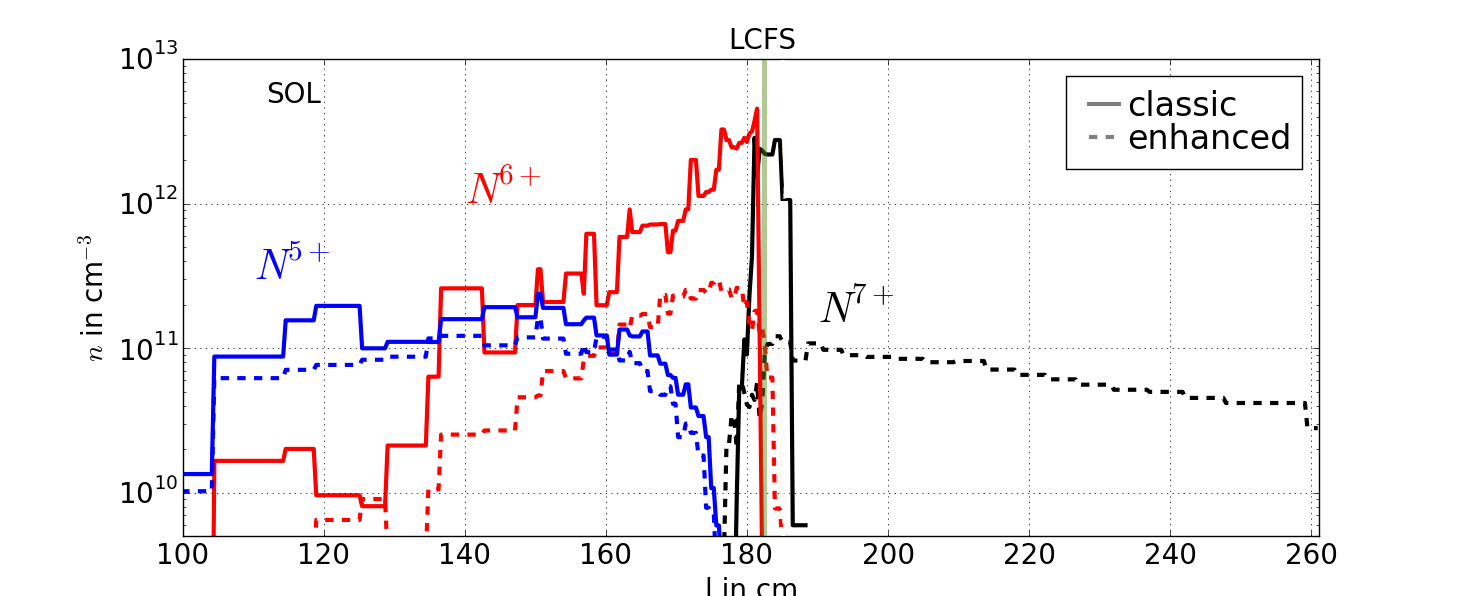}
\caption{Densities of kinetically simulated $N^{5+}$ (blue), $N^{6+}$ (red), and $N^{7+}$ (black) in the puffed ITER baseline scenario as described in the text vs. length along gas puff. Compared are the classical EIRENE physics model (solid) to the one enhanced by drifts, diffusion, and mirror force (dashed).}
\label{fig:diff}
\end{figure}

Fig. \ref{fig:diff} compares the results of this enhanced kinetic ion transport to the aforementioned \emph{classic} EIRENE version including only field-line tracing and energy relaxation. It is constructed in the following way: plotted on the $x$-axis is the length along the gas puff direction (cf. red arrow in Fig. \ref{fig:scheme}), on the $y$-axis is a logarithmic scaling of the particle density in cm$^{-3}$. Shown are the densities of $N^{5+}$ (blue), $N^{6+}$ (red), and $N^{7+}$ (black) obtained using the classic EIRENE ion transport (solid) and the enhanced one (dashed), respecively, while in the latter case we chose $D_\perp=1$ m$^2/$s as the perpendicular diffusion coefficient. The vertical gray line marks the last closed flux surface (LCFS), hence, to the left of it is the scrape-off layer (SOL) and to the right is the confined area.

For the $N^{5+}$ distribution we note almost no difference whether the enhancements in the physical model have been regarded or not, as both treatments give a relatively flat profile at $n_{N^{5+}}\approx10^{11}$ cm$^{-3}$ well in the SOL. For $N^{6+}$, we notice a significant decrease in the peaking at the LCFS by more than an order of magnitude, whereas for $N^{7+}$ we remark a drastic change from a pure peaking at the LCFS in the classic EIRENE treatment towards a smeared out profile well into the confined area in case the drift, diffusion, and mirror force enhancements are regarded. 

While this flattening of the formerly highly peaked profile is mainly due to diffusion, this cannot be established as a general trend. Multiple investigations where either drift or diffusion have been turned off artificially have shown that neither of those two effects is mainly responsible for the simulated plasma profiles but indeed both have to be regarded together.
\section{Conclusions and Outlook}\label{summary}
We presented ITER simulations perturbed by a nitrogen gas puff using kinetic ion transport simulations in EMC3-EIRENE, for the first time applying the physical enhancements in the transport description introduced in \cite{ctpp}. We showed the charge state resolved density distributions in the divertor region, which give important insight on the radiative properties of the plasma and, thus, on the overall confinement performance. Afterwards, we summarized the influence of the external nitrogen gas puff on the main plasma species. Ultimately, we stressed the large influence the physics enhancements of drifts, diffusion, and mirror force have on the simulated profiles.

This case study may serve as a template for ongoing investigations including either different or additional impurity species, e.g. neon or hydrocarbons. EIRENE is able to host for arbitrary species, as long as the necessary reactive data is provided.
\section*{Acknowledgement}
The author thanks P. B\"orner and M. Rack for setting up the simulation case and for fruitful discussions. The author gratefully acknowledges the cooperation with H. Frerichs who provided the initial EMC3-EIRENE ITER baseline geometry. This work has been carried out within the framework of the EUROfusion Consortium and has received funding from the Euratom research and training 
programme 2014-2018 under grant agreement No 633053. This work was supported by the EUROfusion - Theory and Advanced Simulation Coordination (E-TASC) and has received funding from the Euratom research and training programme 2019-2020 under grant agreement No 633053. The views and opinions expressed herein do not necessarily reflect those of the European Commission. The author gratefully acknowledges the computing time granted through JARA-HPC on the supercomputer JURECA at Forschungszentrum J\"ulich.

\bibliographystyle{rackstyleShort}

{\footnotesize

}

\end{document}